\documentclass[preprint,preprintnumbers,amsmath,amssymb]{revtex4}
\usepackage{booktabs}
\usepackage{mathrsfs}
\usepackage{epsfig}
\usepackage{graphicx}% Include figure files
\usepackage{dcolumn}% Align table columns on decimal point
\usepackage{bm}% bold math
\usepackage{amsmath}
\usepackage{slashed}       % slashed p

\begin{document}

\title{The Phenomenological Research on Higgs and dark matter in the Next-to-Minimal Supersymmetric Standard Model}

\author{Zhaoxia Heng, Shenshen Yang, Xingjuan Li, Liangliang Shang$^{[1]}$ 
  \\School of Physics, Henan Normal University, Xinxiang 453007, China
 \vspace{2cm}}

\footnotetext[1]{e-mail: shangliangliang@htu.edu.cn (corresponding author)}

\begin{abstract}
The $Z_3$-invariant next-to-minimal supersymmetric standard model (NMSSM) can provide a candidate for dark matter (DM). It can also be used to explain the hypothesis that the Higgs signal observed on the Large Hadron Collider (LHC) comes from the contribution of the two lightest CP-even Higgs bosons, whose masses are near 125 GeV.  
At present, XENON1T, LUX, and PandaX experiments have imposed very strict restrictions on direct collision cross sections of {dark matter}. 
In this paper, we consider a scenario that the observed Higgs signal is the superposition of two mass-degenerate Higgs in the $Z_3$-invariant NMSSM and scan the seven-dimension parameter space composing of $\lambda, \kappa, \tan\beta, \mu, A_k, A_t, M_1$ via
the Markov chain Monte Carlo (MCMC) method.
We find that the DM relic density, as well as the LHC searches for sparticles, especially the DM direct detections, has provided a strong limit on the parameter space. %Please check intended meaning has been retained.
The allowed parameter space is featured by a relatively small $\mu \le 300$ GeV and about $\tan\beta\in(10,20)$. In addition, the DM is Higgsino-dominated because of $|\frac{2\kappa}{\lambda}|>1$. Moreover, the co-annihilation between $\tilde{\chi}_1^0$ and $\tilde{\chi}_1^\pm$ must be taken into account to obtain the reasonable DM relic density. 
    \end{abstract}

    \maketitle

\section{\label{Introduction}Introduction}
The 125 GeV Higgs boson was discovered in 2012 at the Large Hadron Collider (LHC)~\cite{ATLAS:2012yve,CMS:2012qbp}, which has verified the validity of the standard model (SM) at energy scales around TeV. However, the existence of dark matter (DM) cannot be explained reasonably in the SM, and new physics models beyond the SM are required. The lightest supersymmetric particle (LSP) in various supersymmetric (SUSY) models provide a suitable candidate for the weakly interacting massive particle (WIMP), which is a natural prediction for DM~\cite{Jungman:1995df,Cao:2022chy,Han:2014nba}.

The minimal supersymmetric standard model (MSSM), as one of the most popular new physics models, can provide an elegant solution to the hierarchy problem and predict the lightest neutralino as the DM candidate. 
Although MSSM has remarkable advantages, there are some problems,  such as $\mu$-problem and little hierarchy problem, which has been exacerbated by the LHC experiments in recent years~\cite{Randall:1998uk,Gherghetta:2003he,Wang:2014kja,Wang:2022rfd,Harz:2015qva}. 
These problems can be solved in the next-to-minimal supersymmetric standard model (NMSSM) extending the Higgs sector of MSSM with a gauge singlet field $\hat S$. 
When $\hat S$ develops a vacuum expectation Value (VEV) $v_s$,  an effective $\mu$-term ($\mu_{eff}=\lambda v_s$) is dynamically generated, and its magnitude is naturally at the electroweak scale ~\cite{Ellwanger:2009dp,Maniatis:2009re,Cao:2013gba,Cao:2016uwt}.  
Moreover, the squared mass of SM-like Higgs boson can receive a positive contribution at tree-level because of the interactions among Higgs fields $\lambda \hat S \hat H_u \cdot \hat H_d$ in the NMSSM~\cite{Ellwanger:2011aa,Gunion:2012zd,Kang:2012sy,King:2012is,Cao:2012fz}. 
Furthermore, the mass can be enhanced by singlet-doublet Higgs mixing if the Higgs boson is the next-to-lightest CP-even Higgs state~\cite{Cao:2018rix,Cao:2016cnv,Cao:2016nix}. 
As a result, large radiative corrections to the Higgs boson mass are unnecessary and the little hierarchy problem can be avoided. 
 
Several different methods have been proposed to diagnose whether the discovered 125~GeV Higgs boson
is the superposition of two or more mass-degenerate Higgs signals~\mbox{\cite{Gunion:2012he,Grossman:2013pt,David:2014jla}}.
{The two-Higgs-doublet model (2HDM)}~\cite{Han:2015pwa,Bian:2017gxg,Han:2020ekm,Han:2022juu} and the NMSSM~\cite{Gunion:2012gc, Munir:2013wka, AbdusSalam:2019gnh,Moretti:2015bua,Wang:2015omi,AbdusSalam:2017uzr,Das:2017tob, Shang:2022hbv} have been discussed.
For example, Ref.~\cite{David:2014jla} developed a method testing the presence of multiple Higgs bosons with profile likelihood techniques, which could be
directly used by the ATLAS and CMS collaborations.
It is known that DM direct detection experiments, such as XENON1T~\cite{XENON:2018voc, XENON:2019rxp}, LUX~\cite{LUX:2017ree} and PandaX~\cite{PandaX-II:2017hlx}, have imposed strict limits on DM~\cite{Wang:2021oha,Wang:2021nbf}.
Consequently, we will scan the parameter space and explore the phenomenology considering the latest DM experiments in two mass-degenerate 125 GeV Higgs bosons of the $Z_3$-invariant NMSSM. Note that we let the mass of Bino $M_1$ be free, which could change the composition of DM and is different from our previous work~\cite{Shang:2020uog}.

This paper is arranged as follows:  in Section \ref{sec:model}, we briefly introduce the $Z_3$-invariant NMSSM and explain our scanning strategy. In Section \ref{sec:constrain}, we show properties of DM confronted with DM relic density and direct detection experimental results in two mass-degenerate 125 GeV Higgs bosons scenarios. In {Section} %MDPI: The secton 5 give a summary, please check if the section number is wrong here, please check all section numbers in this paragraph.
 \ref{sec:sum}, we provide a summary.

\section{\label{sec:model}Model and Scan Strategy}
    \subsection{\label{model}Basic of the $Z_3$-Invariant NMSSM}

The superpotential in the $Z_3$-invariant NMSSM consists of the Yukawa term  $W_{F}$ in the MSSM and terms that are related to the additional gauge singlet chiral superfield $\hat{S}$: 
\begin{eqnarray}
%\begin{split}
   W&=&W_{F}+\lambda \hat{H}_{u}\cdot \hat{H}_{d} \hat{S} + \frac{1}{3}\kappa\hat{S}^3
%\end{split}
\end{eqnarray}
where the parameters $\lambda$ and $\kappa$ are dimensionless and there is no $\mu$-term in $W_{F}$. At the tree-level, the Higgs scalar potential $V$ can be deduced from the superpotential $W$~\cite{Ellwanger:2009dp}:
\begin{eqnarray}
%\begin{split}
   V&=&V_{F}+V_{D}+V_{\rm{soft}}
%\end{split}
\end{eqnarray}
\begin{eqnarray}
   V_{F}&=&|\lambda S|^2 (|H_u|^2 + |H_d|^2) +
   |\lambda H_u \cdot H_d + \kappa S^2|^2 \nonumber \\
   V_{D}&=&\frac{1}{8} ( {g_1}^2 + {g_2}^2) (|H_d|^2 - |H_u|^2)^2 + \frac{1}{2} {g_2}^2 | H_u^{\dagger} \cdot H_d |^2  \\
   V_{\rm{soft}} &=& M_{H_u}^2 |H_u|^2 + M_{H_d}^2 |H_d|^2 + M_S^2 |S|^2 + (\lambda A_{\lambda} S H_u \cdot H_d + \frac{1}{3} \kappa A_{\kappa} S^3 + h.c.), \nonumber   
    \end{eqnarray}
where $A_{\lambda, \kappa}$ are the soft SUSY breaking trilinear parameters, and $g_1$ and $g_2$ are the gauge couplings in the $U(1)_Y$ and $SU(2)_L$ gauge symmetries. Considering the minimization conditions of the Higgs potential after electroweak symmetry breaking, $M_{H_d,H_u,S}^2$ are substituted by their vacuum expect values $\langle H_u \rangle=v_u$, $\langle H_d \rangle=v_d$ and $\langle H_s \rangle=v_s$. 
Then, an effective $\mu$-term is generated as $\mu=\lambda v_s$, allowing the $\mu$-problem in the MSSM to be solved~\cite{Maniatis:2009re}. 
{{It is known} %MDPI: We removed the bold. Please confirm this revision.
 that $\mu \le 300$ GeV is important for electroweak symmetry breaking in the NMSSM because $\mu = \lambda v_s$ and $v_s$ should be near the electroweak scale for the singlet generally to have critical effects on electroweak phase transition~\cite{Kozaczuk:2014kva,Kozaczuk:2013fga}.}
Finally, there are six independent parameters left in the Higgs sector of the NMSSM at the tree-level: 
    \begin{equation}
        \lambda, \quad \kappa, \quad \tan{\beta} = \frac{v_u}{v_d}, \quad \mu, \quad A_{\lambda}, \quad A_{\kappa}
        \label{eq: param}        
    \end{equation}
%    $M_A=\frac{2\mu(A_{\lambda}+\kappa v_s)}{\sin{2\beta}}$

In the $Z_3$-invariant NMSSM, it is convenient to use the following definition:
        \begin{eqnarray}
        h_0 = \cos{\beta} H_u+\varepsilon \sin{\beta} H_d^* , \ H_0 = \sin{\beta} H_u+\varepsilon \cos{\beta} H_d^* 
            \end{eqnarray}
where  $\varepsilon$  is 2-dimensional antisymmetric tensor and $\varepsilon_{12} =-\varepsilon_{21}=1,   \varepsilon_{11}= \varepsilon_{22}=0$~\cite{Miller:2003ay}. Now, the $h_i$ = ($h_0$, $H_0$, $S$)$^T$ can be written as
    \begin{eqnarray}
        h_0 =  \left ( \begin{array}{c}
                    H^+ \\
                    \frac{S_1 + \mathrm{i} P_1}{\sqrt{2}}
                \end{array} \right),~~
        H_0 = \left ( \begin{array}{c}
                    G^+ \\
                    v + \frac{S_2 + \mathrm{i} G^0}{\sqrt{2}}
                \end{array} \right),~~
        S  =  v_s +\frac{1}{\sqrt{2}} \left( S_3 + \mathrm{i} P_2 \right),
    \end{eqnarray}
where $v^2 = v_u^2 + v_d^2$,  $G^+$ and $G^0$ are the Goldstone bosons. The above equation manifests that $H_0$ corresponds to the Higgs field in the SM. The  CP-even Higgs mass matrix in the basis ($S_1$, $S_2$, $S_3$) at tree-level can be described as
    \begin{eqnarray}
        M_{S_1S_1}^2 &=&  M^2_A + (M^2_Z -\lambda^2 v^2) \sin^2 2\beta, \quad M_{S_1S_2}^2 =  -\frac{1}{2}(M^2_Z-\lambda^2 v^2)\sin4\beta, \nonumber \\
        M_{S_1S_3}^2 &=&  -(\frac{M^2_A}{2\mu/\sin2\beta}+\kappa v_s) \lambda v\cos2\beta, \quad M_{S_2S_2}^2 =  M_Z^2\cos^2 2\beta +\lambda^2v^2\sin^2 2\beta, \nonumber \\
        M_{S_2S_3}^2 &=&  2\lambda\mu v[1-(\frac{M_A}{2\mu/\sin2\beta})^2 -\frac{\kappa}{2\lambda}\sin2\beta], \nonumber \\
        M_{S_3S_3}^2 &=&  \frac{1}{4}\lambda^2 v^2(\frac{M_A}{\mu/\sin2\beta})^2 +\kappa v_s A_{\kappa}+4(\kappa v_s)^2 -\frac{1}{2}\lambda\kappa v^2 \sin 2\beta, \nonumber
    \end{eqnarray}
where $M_A^2=2\mu(A_\lambda+\kappa v_s)/\sin2\beta$. With the rotation matrix $U$, we can diagonalize the mass matrix $M^2$ and obtain the physical mass eigenstates $H_{i} = \sum\limits_{j=1}^3 U_{ij} S_j$. In addition, the CP-odd mass eigenstates $A_1$ and $A_2$ can be derived in the same way. We assume $M_{H_1}<M_{H_2}<M_{H_3}$ and $M_{A_1} < M_{A_2}$. If the main component of $H_{i}$ is the $S_2$ field, $H_{i}$ is called the SM-like Higgs (denoted by $h$). 
Compared to the case in the MSSM, the mass of SM-like Higgs in the NMSSM at the tree-level could be enhanced because of the additional term $\lambda^2v^2\sin^2 2\beta$ and the mixing effect of $(S_2, S_3)$ when $M_{S_3S_3}^2 < M_{S_2S_2}^2$. Therefore, it needs less radiative corrections in the NMSSM to obtain the 125 GeV SM-like Higgs compared with that in the MSSM  ~\cite{King:2012tr,Jeong:2012ma,Badziak:2013bda}. 
{The observable} %MDPI: We removed the bold. Please confirm this revision.
 $O_{if}$ can be used to explain that how it is possible to have multi mass-degenerate Higgs bosons
under the present measurements of the Higgs boson properties at the LHC~\cite{Shang:2022hbv},
\begin{equation}
	\label{R}
	\begin{split}
		O_{i f} &= \sum_{\alpha} O^{\alpha}_{i f} \\
		O^{\alpha}_{i f} & =\sigma^{H_\alpha}_{i} B^{H_\alpha}_{f}
	\end{split}
\end{equation}
where $i$ denotes the production modes and $f$ denotes the decay modes. The major $O_{if}$ are listed in Table 1 in Ref.~\cite{Shang:2022hbv}, 
of which best-fit values and uncertainties can be found in Refs.~\cite{ATLAS:2016neq,ATLAS:2015zhl,CMS:2018uag,ATLAS:2019nkf,ATLAS:2020evk}.
Note that the index $\alpha$ of the resonance should be summed over in Equation~(\ref{R}) if there are two or more mass-degenerate bosons.

 The masses of charged Higgs bosons $H^\pm$  at tree-level  are given by
    \begin{eqnarray}
      M^2_{H^\pm} &=& M^2_A+M^2_W-\lambda^2v^2
    \end{eqnarray} 

 The neutralinos in the NMSSM are the mixtures of the fields Bino $\tilde{B}^{0}$, Wino $\tilde{W}^{0}$, 
Higgsinos $\tilde{H}_{d,u}^{0}$, and Singlino $\tilde{S}^{0}$.  In the basis $\psi^{0}=\left(-i \tilde{B}^{0},-i \tilde{W}^{0}, \tilde{H}_{d}^{0}, \tilde{H}_{u}^{0}, \tilde{S}^{0}\right)$, one can obtain the symmetric neutralino mass matrix as
\begin{equation}\mathcal{M}_{0}=\left(\begin{array}{ccccc}
M_{1} & 0 & -\frac{g_{1} v_{d}}{\sqrt{2}} & \frac{g_{1} v_{u}}{\sqrt{2}} & 0 \\
& M_{2} & \frac{g_{2} v_{d}}{\sqrt{2}} & -\frac{g_{2} v_{u}}{\sqrt{2}} & 0 \\
& & 0 & -\mu & -\lambda v_{u} \\
& & & 0 & -\lambda v_{d} \\
& & & & \frac{2 \kappa}{\lambda} \mu
\end{array}\right)
\label{e:dm_cp}
\end{equation}
where $M_1$ and $M_2$ denote the gaugino soft breaking masses. With the unitary rotation matrix $N$, one can diagonalize the mass matrix  $\mathcal{M}_{0}$ to obtain the mass eigenstates
 \mbox{$\tilde{\chi}_{i}^{0}=N_{i j} \psi_{j}^{0}$ (${i} ,{j}=1,2,3,4,5$)} and the mass eigenstates labeled in mass-ascending order. The lightest supersymmetric particle (LSP) $\tilde{\chi}_{1}^{0}$ can be regarded as one of the DM candidates.
 
  Analogously,  in the gauge-eigenstate basis 
$\psi^{\pm}=\left(\tilde{W}^{+}, \tilde{H}_{u}^{+}, \tilde{W}^{-}, \tilde{H}_{d}^{-}\right)$, the chargino mass matrix can be given by
\begin{equation}M_{\chi^{\pm}}=\left(\begin{array}{cc}
0 & X^{\mathrm{T}} \\
X & 0
\end{array}\right), \quad X=\left(\begin{array}{cc}
M_{2} & \sqrt{2} s_{\beta} M_{W} \\
\sqrt{2} c_{\beta} M_{W} & \mu 
\end{array}\right)\end{equation}

One can obtain the mass eigenstates by two unitary rotation matrices as follows: 
\begin{equation}\left(\begin{array}{c}
\chi_{1}^{+} \\
\chi_{2}^{+}
\end{array}\right)=U^{+}\left(\begin{array}{c}
\tilde{W}^{+} \\
\tilde{H}_{u}^{+}
\end{array}\right), \quad\left(\begin{array}{c}
\chi_{1}^{-} \\
\chi_{2}^{-}
\end{array}\right)=U^{-}\left(\begin{array}{c}
\tilde{W}^{-} \\
\tilde{H}_{d}^{-}
\end{array}\right)\end{equation}
\begin{equation}\operatorname{diag}\left(M_{\chi_{1}^{\pm}}, M_{\chi_{2}^{\pm}}\right)=\left(U^{+}\right)^{*} X\left(U^{-}\right)^{\dagger}\end{equation}

\subsection{\label{sec:scan}Scan Strategies and Constraints on the Parameter Space of NMSSM}
$A_\lambda$ is fixed at 2 TeV because the masses of charged Higgs bosons are usually large considering the constraints from the LHC and $M_{H^\pm}$ are determined by the parameter $A_\lambda$ as shown in Equation (7). In addition, the soft breaking parameters except $A_t$ in the slepton and squark sectors are fixed at 2 TeV because the stop trilinear coupling $A_t$ plays an significant role in the 125 GeV Higgs boson via loop-corrected contributions. 
{{Moreover,} %MDPI: We removed the bold. Please confirm this revision.
 the Wino mass $M_2$ is fixed at 2 TeV for simplicity because the wino-dominated DM could hardly satisfy limits from both DM and LHC experiments~\cite{Cao:2016nix}.} 
As a result, the Markov chain Monte Carlo (MCMC) scan is utilized in these parameters, 
     \begin{equation}\label{eq:input}
    \begin{split}
        &0<\lambda<0.75,\quad \left| \kappa \right|<0.75,\quad 1<\tan{\beta}<60,\quad -1~{\rm TeV}\leq M_1 \leq 1~{\rm TeV},\\
        &100~{\rm GeV}\leq \mu \leq 1~{\rm TeV},\quad \left|A_\kappa\right| \leq 1~{\rm TeV},\quad \left|A_t\right| \leq 5~{\rm TeV}.
    \end{split}
    \end{equation}
 
During the scan, we select the samples that are consistent with these constraints,
\begin{itemize}
\item  All of the constraints are implemented in the package \texttt{NMSSMTools}-5.5.3 ~\cite{Ellwanger:2004xm,Ellwanger:2005dv}, which includes the Z-boson invisible decay, the LEP search for sparticles (i.e., the lower bounds on various sparticle masses and the upper bounds on the chargino/neutralino pair production rates), the B-physics observables such as the branching ratios for $B \to X_s \gamma$ and $B_s \to \mu^+ \mu^-$, and the discrepancy of the muon anomalous magnetic moment. The latest measured results are utilized for certain observables with an experimental central value, and the selected samples could explain these results at 2$\sigma$~level.

\item Constraints on the direct searches for Higgs bosons at LEP, Tevatron, and LHC. These constraints are implemented through the packages \texttt{HiggsSignals}~\cite{Bechtle:2013xfa,Bechtle:2014ewa,Stal:2013hwa} for 125~GeV Higgs data fit and \texttt{HiggsBounds}~\cite{Bechtle:2008jh,Bechtle:2011sb} for non-standard Higgs boson
search at colliders. Two nearly mass-degenerate CP-even Higgs bosons with masses 122 GeV $\le M_{h_1,h_2} \le$ 128 GeV are required.

\item  The package \texttt{micrOMEGAs} ~\cite{Belanger:2008sj,Belanger:2010gh} embedded in \texttt{NMSSMTools} is utilized to calculate the thermally averaged cross section, the DM relic density, and the spin-dependent (SD) and spin-independent (SI) DM-nucleon cross sections of DM. 
{{The} %MDPI: We removed the bold. Please confirm this revision.
 LSP $\tilde{\chi}_1^0$ should be with a thermal abundance matching the observed DM density. What's more,
DM could be composed of a lightest neutralino, an axion~\cite{Preskill:1982cy} or gravitino~\cite{Ellis:1983ew}, so that we suppose that there was a large amount of DM in the early universe, and they reached the current Planck observation $\Omega_{\text{DM}}h^2 = 0.120 \pm 0.01$ as they freezed out~\cite{ParticleDataGroup:2018ovx,WMAP:2012nax,Planck:2013pxb}.
Consequently, the DM relic density is required to be less than the central value 0.12 in our work.
In addition to the relic density, the DM should be compatible with direct detection rates in accordance with current limits, which come from LUX-2017~\cite{LUX:2017ree},  XENON1T-2019  ~\cite{XENON:2019rxp} for SD cross sections, and XENON1T-2018 ~\cite{XENON:2018voc} for SI cross sections. It is noticed that the DM-nucleon cross sections should be scaled by a factor $\Omega h^2 / 0.120$ given that the LSP $\tilde{\chi}_1^0$ is only one of the DM candidates.}

\item {{Results} %MDPI: We removed the bold. Please confirm this revision.
 from LHC searching sparticles. 
Processes $pp \to \tilde{\chi}_1^\pm \tilde{\chi}_1^0$, $pp \to \tilde{\chi}_1^\pm \tilde{\chi}_2^0$ and $pp \to \tilde{\chi}_1^+ \tilde{\chi}_1^-$ are put into Prospino2~\cite{Beenakker:1996ed} to calculate their NLO cross sections at LHC 13 TeV. Then, these processes and cross sections are fed into SModelS-2.1.1~\cite{Kraml:2013mwa}, which decomposes spectrums and converts them into simplified model topologies to compare with the results interpreted from the LHC.}

\end{itemize}

\section{\label{sec:constrain} Properties of DM}

\begin{figure}[htbp]
\includegraphics[width=10cm]{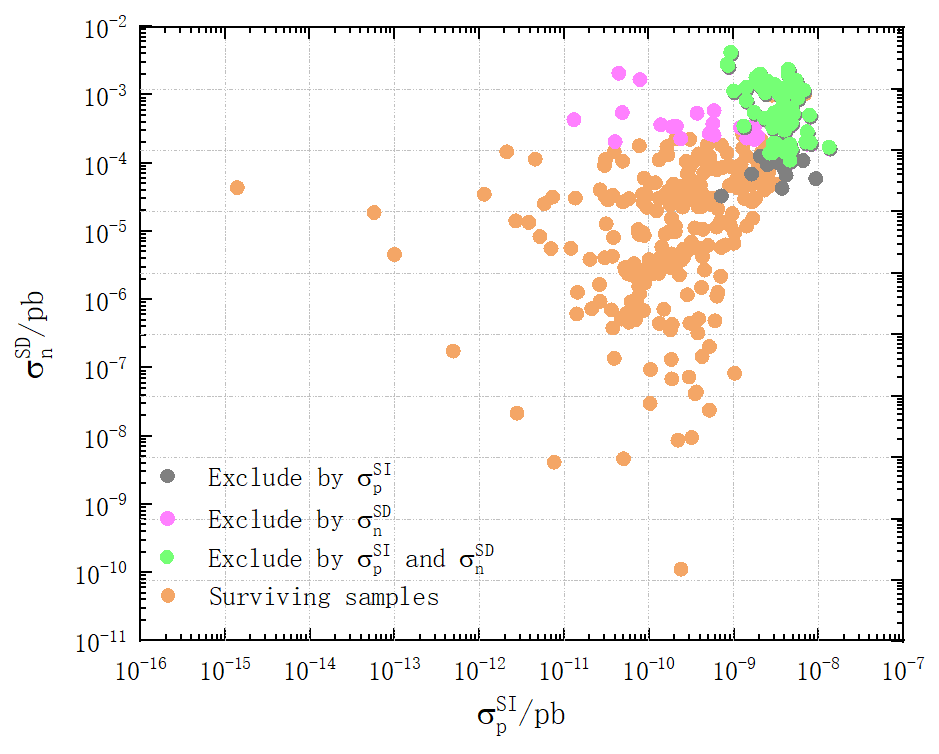} 
\caption{{Spin-dependent (SD) and spin-independent (SI)} %MDPI: Please use the scientific notation, e.g., "$8 \times 10^{3}$", not "8E3". We corrected them. Please confirm.
 cross sections for samples satisfying the constraints listed in {Section} %MDPI: Please check the section number
 \ref{sec:model}. 
The limitation comes from LUX-2017~\cite{LUX:2017ree},  XENON1T-2019~\cite{XENON:2019rxp} for SD cross sections, and XENON1T-2018~\cite{XENON:2018voc} for SI cross sections. }  %MDPI: Please check the section number
\label{fig1}
\end{figure}

We calculate the SI and SD cross sections for samples satisfying the constraints listed in {Section} %MDPI: Please check if the section number is right
 \ref{sec:model}.  Additionally, we project these samples in the plane of  $\sigma_p^{SI}-\sigma_n^{SD}$ in Figure~\ref{fig1}, where gray samples are excluded by the SI detection coming from XENON1T-2018~\cite{XENON:2018voc},  XENON1T-2019~\cite{XENON:2019rxp}, pink samples are excluded by the SD detection coming from LUX-2017~\cite{LUX:2017ree}, and orange (green) samples are consistent with (excluded by) both the SD and SI detections.
From this figure, we find that the DM direct detections impose a strong constraint on the parameter space in the scenario of two mass-degenerate Higgs bosons in the $Z_3$-invariant NMSSM, and
the constraints from direct and indirect detections of DM are complementary.

\begin{figure}[htbp]
\includegraphics[width=6.5cm]{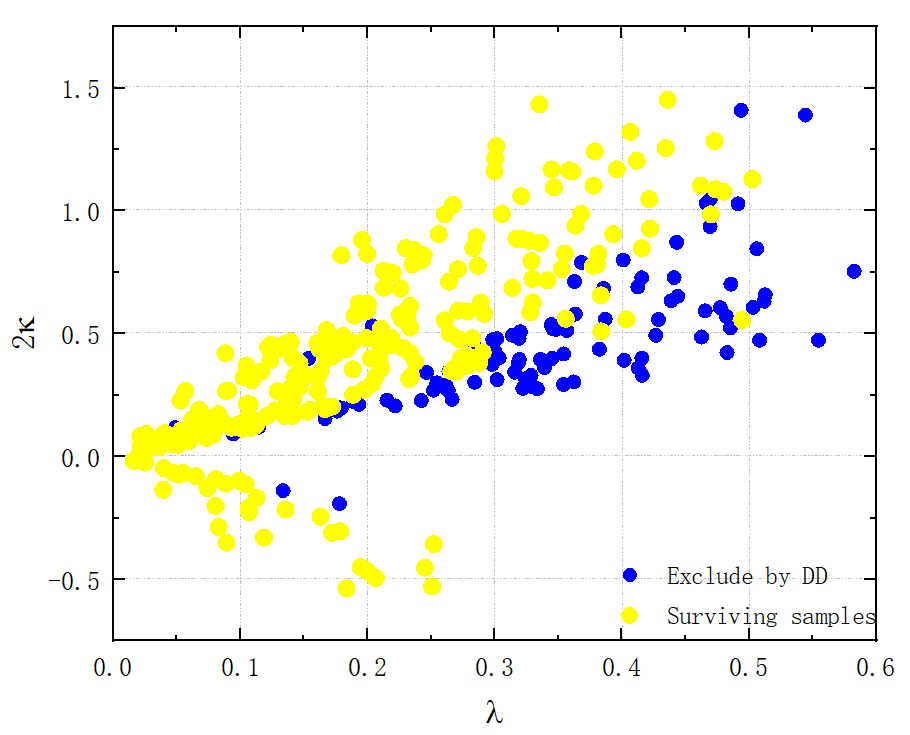}
\includegraphics[width=6.5cm]{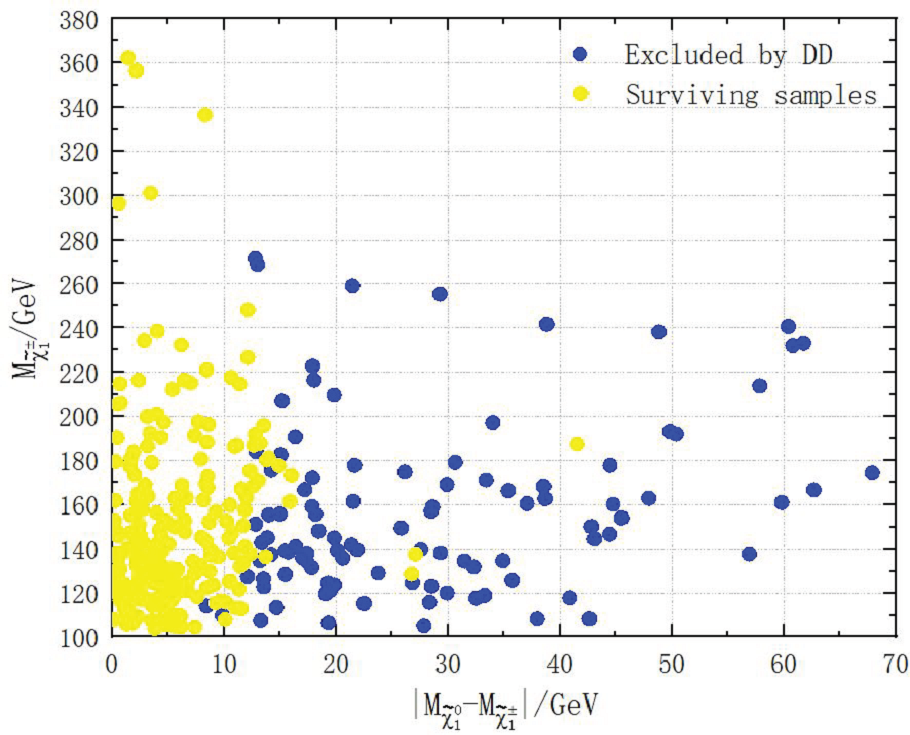}
\caption{Samples projected on the left plane of $\lambda - 2\kappa$ and right plane of $|M_{\tilde{\chi}_1^\pm} - M_{\tilde{\chi}^0_1}|-M_{\tilde{\chi}_1^\pm}$. Blue samples are excluded by the DM direct detections but yellow samples are consistent with these~detections.}
\label{fig2}
\end{figure}

We project the samples on the left plane of $\lambda - 2\kappa$ and right plane of $|M_{\tilde{\chi}_1^\pm} - M_{\tilde{\chi}^0_1}|-M_{\tilde{\chi}_1^\pm}$ in Figure~\ref{fig2}. 
From the left plane, {{we can} %MDPI: We removed the bold. Please confirm this revision.
 see that most of the surviving samples fall within the range of $|\frac{2\kappa}{\lambda}|>1$, which leads to decoupling Singlino in DM composition, as shown in Equation (\ref{e:dm_cp}}). From the right plane, we can see that 
samples with large mass differences between $\tilde{\chi}_1^0$ and  $\tilde{\chi}_1^\pm$ are almost ruled out by DM direct detections because the co-annihilation between $\tilde{\chi}_1^0$ and $\tilde{\chi}_1^\pm$ must be taken into account to obtain the reasonable DM relic density.
Masses of $\tilde{\chi}_1^0$ and  $\tilde{\chi}_1^\pm$ are nearly close to each other within $10\%$ in the range from about 96 GeV to 240 GeV. 
{{In addition,} %MDPI: We removed the bold. Please confirm this revision.
 we find that 
samples being consistent with DM direct detections are featured by a relatively small $\mu \le 300$ GeV, with the value of $\tan\beta$ between about 10 and 20.}

\begin{figure}[htbp]
\includegraphics[width=6.7cm]{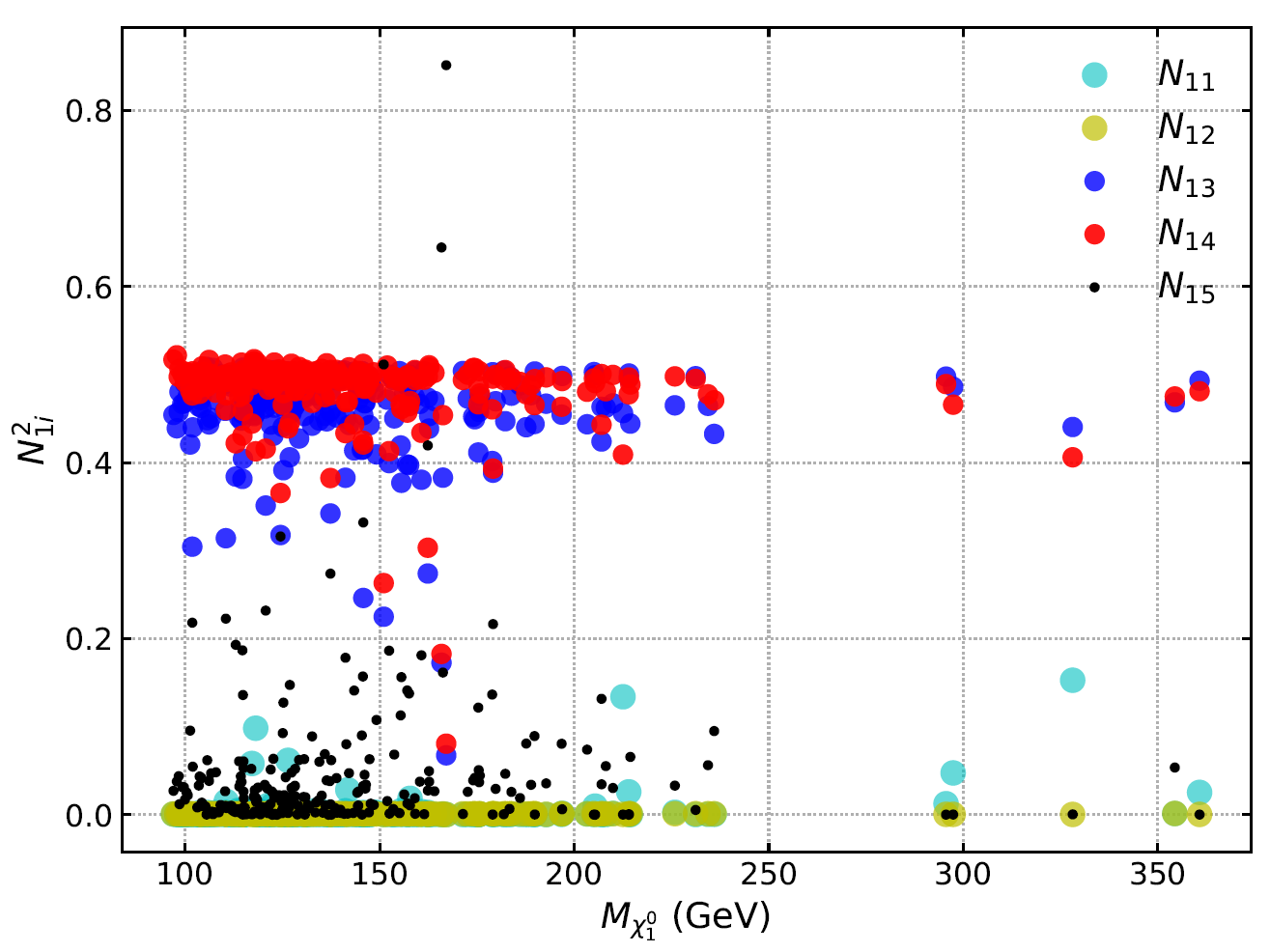}
\includegraphics[width=6.7cm]{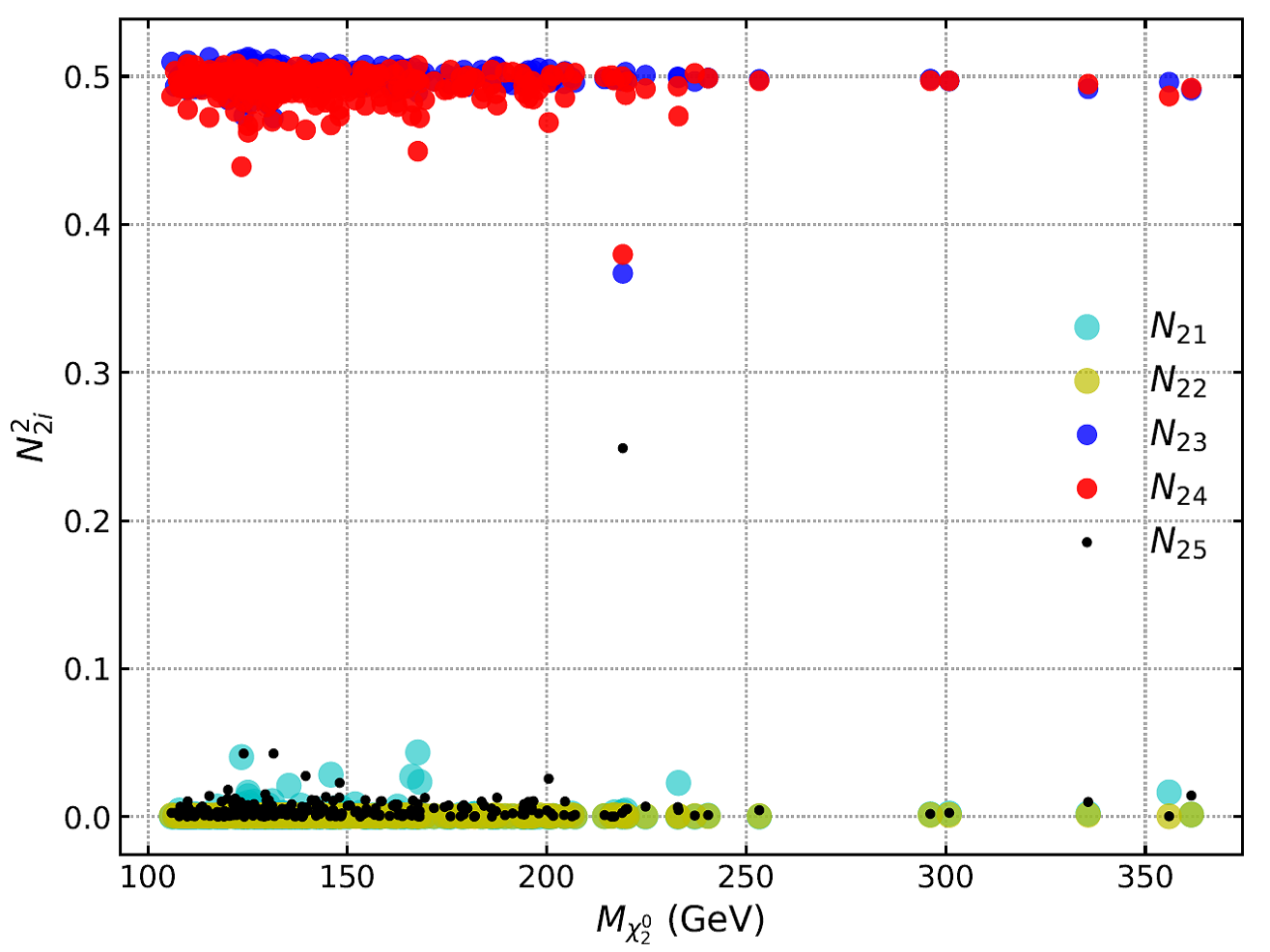}\\
\includegraphics[width=6.7cm]{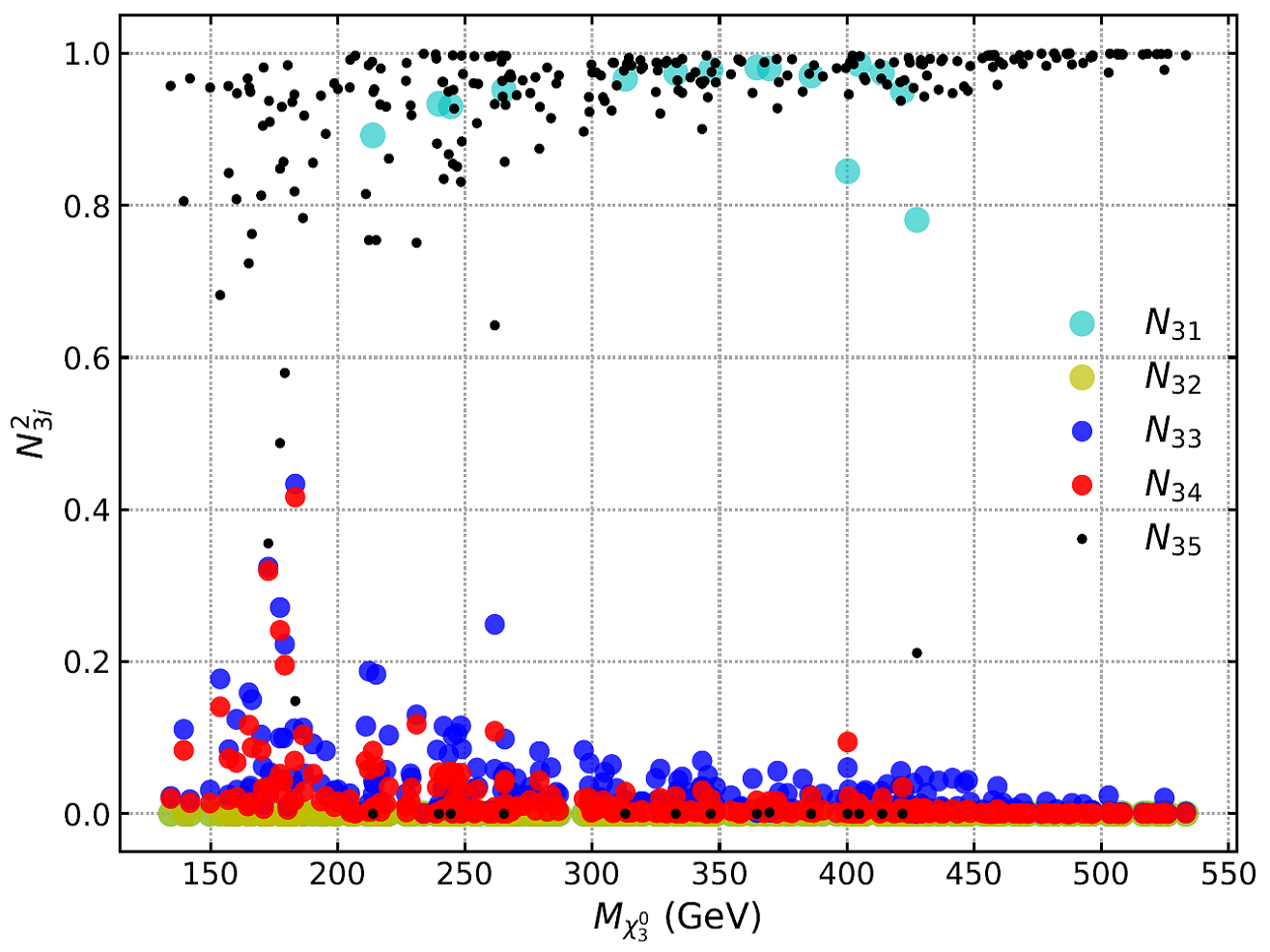}
\caption{Compositions of $\tilde{\chi}_1^0$,  $\tilde{\chi}_2^0$, and  $\tilde{\chi}_3^0$ for samples being consistent with DM direct detections.}
\label{fig-comp}
\end{figure}

We show compositions of $\tilde{\chi}_1^0$,  $\tilde{\chi}_2^0$, and  $\tilde{\chi}_3^0$ for samples being consistent with DM direct detections Figure~\ref{fig-comp}. From this figure, {{we} %MDPI: We removed the bold. Please confirm this revision.
 can see that $\tilde{\chi}_1^0$ and  $\tilde{\chi}_2^0$ are Higgsino-dominated, and $\tilde{H}_{d}^{0}$ and $\tilde{H}_{u}^{0}$ components are comparable; however, $\tilde{\chi}_3^0$ is Singlino-dominated}.

\begin{figure}[htbp]
	\includegraphics[width=16cm]{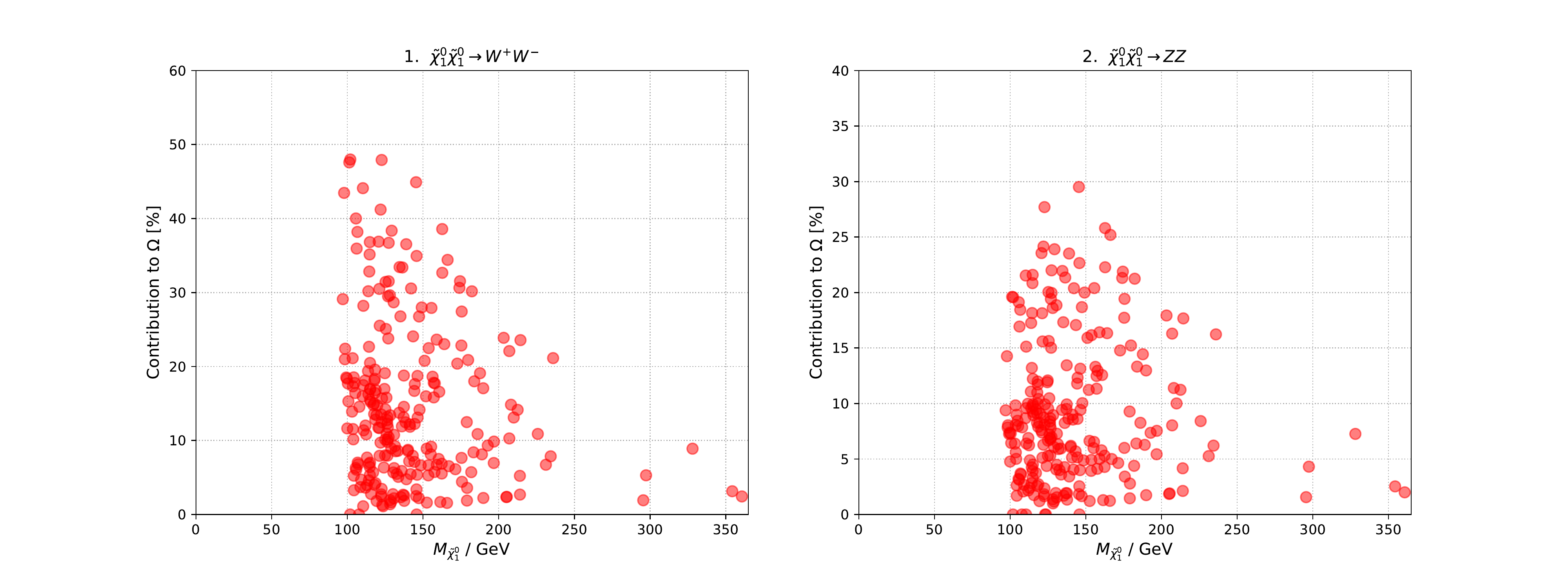}
	\includegraphics[width=16cm]{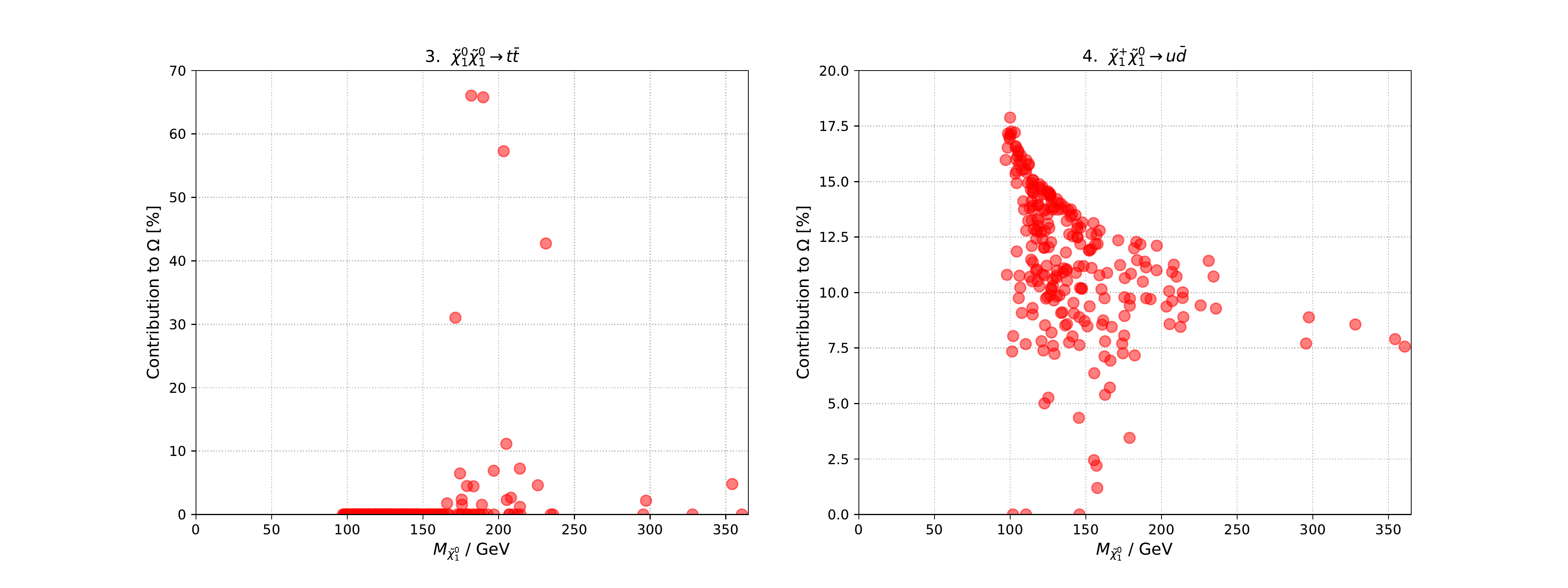}
	\includegraphics[width=16cm]{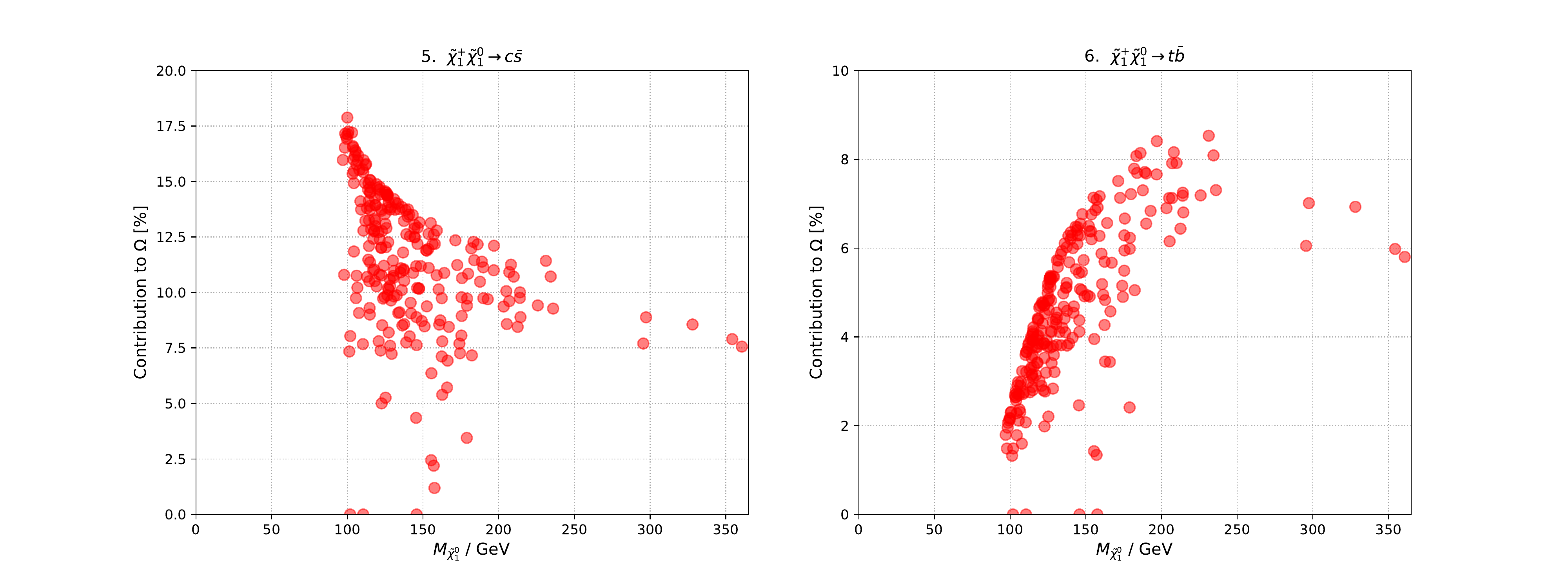}
	\includegraphics[width=16cm]{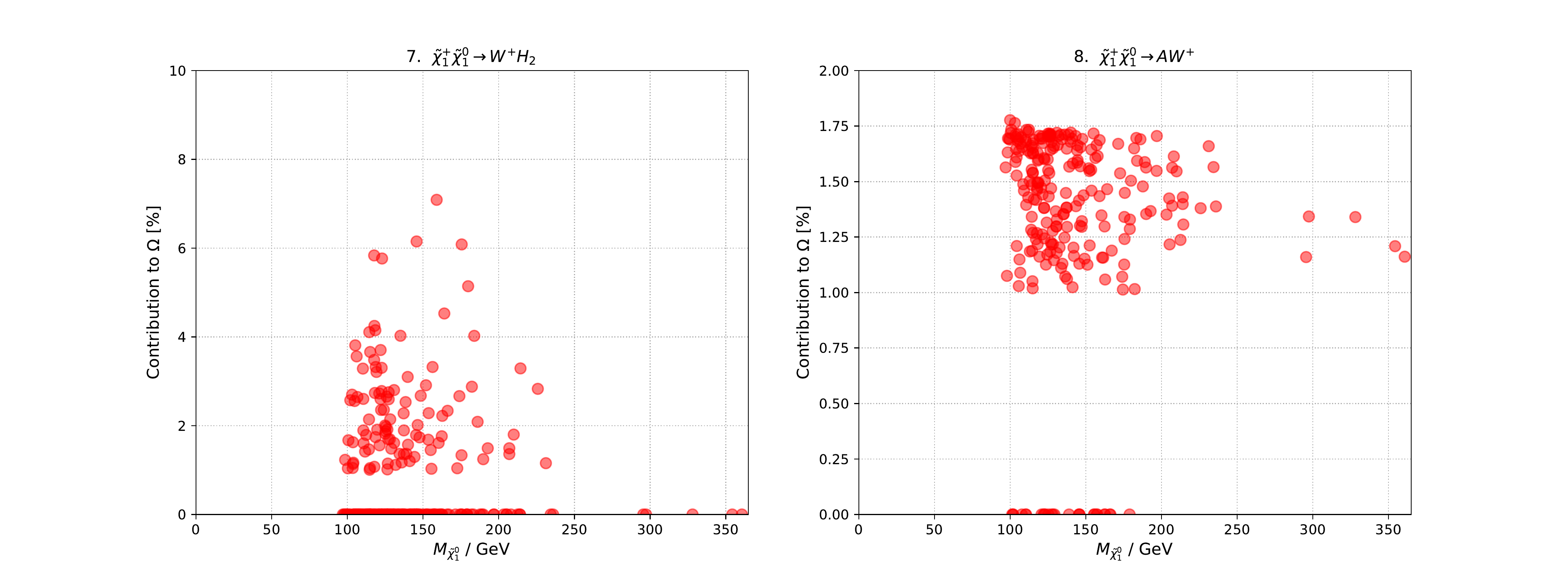}

	\caption{Contributions of annihilation and co-annihilation processes to DM relic density for samples consistent with DM direct detections. Note that samples with zero percents make less of a contribution than 1\% because that their exact values are not output via micrOMEGAs.}
	\label{fig5}
\end{figure}

\begin{figure}[htbp]

%\begin{adjustwidth}{-\extralength}{0cm}
\centering %% If there is a figure in wide page, please release command \centering
\includegraphics[width=16cm]{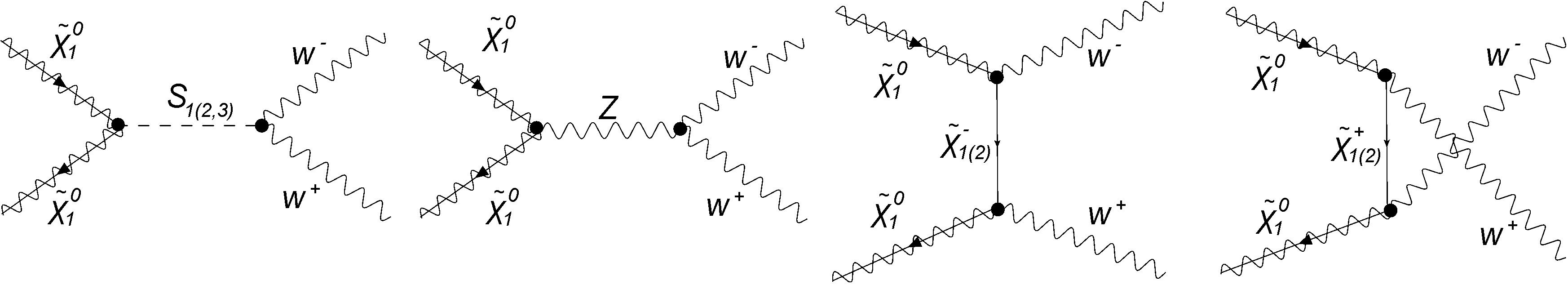}
\includegraphics[width=14cm]{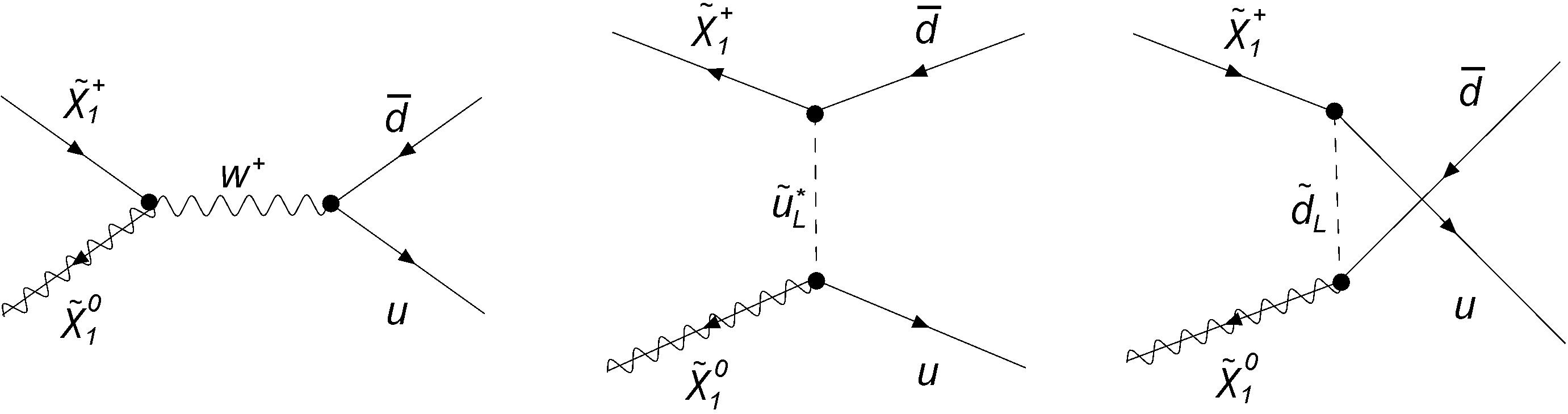}
%\end{adjustwidth}
\caption{Typical tree-level diagrams contributing to DM annihilation in the first row and co-annihilation in the second row.} %Please check intended meaning has been retained.
\label{tree}
\end{figure}

We show contributions to DM relic density for annihilation and co-annihilation modes for samples being consistent with DM direct detections in Figure~\ref{fig5}. 
The primary annihilation channel is $ \tilde{\chi}_{1}^{0} \tilde{\chi}_{1}^{0} \rightarrow W^{+} W^{-}$, as shown in the first row, of which tree-level Feynman diagrams are shown in the first row in Figure~\ref{tree}. 
Its distribution is approximately triangular with few zero points. The LSP mass ranges from about 97 GeV to 361 GeV, of which contributions are about 29.1$\%$ and 2.44$\%$, respectively. 
The largest contribution is about $48.0 \%$ at $M_{\tilde{\chi}_{1}^{0}}=102\, \mathrm{GeV}$, and the smallest contribution is about $1.11 \%$ at $M_{\tilde{\chi}_{1}^{0}}=124\, \mathrm{GeV}$.

The sub-dominant annihilation channel {is} $ \tilde{\chi}_{1}^{0} \tilde{\chi}_{1}^{0} \rightarrow Z  Z$,
as shown in the first row. Its distribution resembles that of the primary annihilation with the exception of a lower peak. The largest contribution is about $29.5 \%$ at $M_{\tilde{\chi}_{1}^{0}}=145\, \mathrm{GeV}$. 
However, the annihilation channel $ \tilde{\chi}_{1}^{0} \tilde{\chi}_{1}^{0} \rightarrow t \bar{t}$ in the second row is not as important as those we just mentioned above. 
Its contribution is sparse, with many zero percents, and ranges from about $1.21 \%$ to $66.0\%$.

The co-annihilation between $\tilde{\chi}_1^0$ and $\tilde{\chi}_1^\pm$ must be taken into account to obtain the reasonable DM relic density.
The primary channels in the co-annihilation mode are $ \tilde{\chi}_{1}^{+} \tilde{\chi}_{1}^{0} \rightarrow u  \bar{d}$ and $\tilde{\chi}_{1}^{+} \tilde{\chi}_{1}^{0} \rightarrow c\bar{s}$, as shown in the second row and the third row in Figure~\ref{fig5}, respectively.
The tree-level Feynman diagrams for $ \tilde{\chi}_{1}^{+} \tilde{\chi}_{1}^{0} \rightarrow u  \bar{d}$ are shown in the second row in Figure~\ref{tree}. 
The LSP mass ranges from about 97 GeV to 361 GeV for both channels, of which contributions are about 16.0$\%$ and 7.56$\%$, respectively,
but the majority of the samples are {located at about $M_{\tilde{\chi}_{1}^{0}} \in [97, 236]$ GeV with contribution percents $[16.0, 9.23]\%$.} In addition,
the largest contribution {is $17.9 \%$ at $M_{\tilde{\chi}_{1}^{0}}=100 \,\mathrm{GeV}$}, and the smallest contribution is $1.20 \%$ at $M_{\tilde{\chi}_{1}^{0}}=156\, \mathrm{GeV}$. 
The sub-dominant co-annihilation channels are
$ \tilde{\chi}_{1}^{+} \tilde{\chi}_{1}^{0} \rightarrow t  \bar{b}$ and $ \tilde{\chi}_{1}^{+} \tilde{\chi}_{1}^{0} \rightarrow W^{+} h_{2}$, as shown in the third row and the last row in Figure~\ref{fig5}.
The distribution of $ \tilde{\chi}_{1}^{+} \tilde{\chi}_{1}^{0} \rightarrow t  \bar{b}$ shows an opposite trend compared to that of $\tilde{\chi}_{1}^{+} \tilde{\chi}_{1}^{0} \rightarrow u  \bar{d}$ or $ \tilde{\chi}_{1}^{+} \tilde{\chi}_{1}^{0} \rightarrow c\bar{s}$ {in the region \mbox{$M_{\tilde{\chi}_{1}^{0}} \in [100, 150] $~GeV.}}
The largest contribution for $ \tilde{\chi}_{1}^{+} \tilde{\chi}_{1}^{0} \rightarrow t  \bar{b}$ is about $8.53 \%$ at \mbox{$M_{\tilde{\chi}_{1}^{0}}= 231~\mathrm{GeV}$.}
The largest contribution for $ \tilde{\chi}_{1}^{+} \tilde{\chi}_{1}^{0} \rightarrow W^{+} h_{2}$ is about $7.09 \%$ at $M_{\tilde{\chi}_{1}^{0}}=159~\mathrm{GeV}$.        
Finally, the last co-annihilation channel is $ \tilde{\chi}_{1}^{+} \tilde{\chi}_{1}^{0} \rightarrow A W^{+}$, as shown in the last row in Figure~\ref{fig5}, of which contributions do not exceed $2.0 \%$.
        
In conclusion, we find that the LSP annihilation makes a contribution to the DM relic density in the allowed parameter space, but its contribution is insufficient to obtain the proper density. To achieve the observed value, the major LSP co-annihilation with $\tilde{\chi}_1^\pm$ should be considered. 

\section{\label{sec:sum}Summary}

In this paper, we 
study the property of the allowed parameter space in the $Z_3$-invariant-NMSSM. We consider two mass-degenerate Higgs bosons as the observed 125 GeV Higgs, LHC searches for sparticles, the DM relic density, and the DM direct detections. These detections come from
LUX-2017 and XENON1T-2019 for the SD cross sections, and XENON1T-2018 for the SI cross sections.
We perform the MCMC scan over the seven-dimension parameter space composed of $\lambda, \kappa, \tan\beta, \mu, A_k, A_t, M_1$. 

Our study indicates that there are still samples capable of predicting the observed 125 GeV Higgs in the case of two mass-degenerate neutral CP-even Higgs bosons in the $Z_3$-invariant-NMSSM. However, the DM relic density, and the LHC searches for sparticles, especially the DM direct detections, have provided a strong limit on the parameter space. 
The allowed parameter space is featured by a relatively small $\mu \le 300$ GeV and about $\tan\beta\in(10,20)$. In addition, the DM is Higgsino-dominated because of $|\frac{2\kappa}{\lambda}|>1$. Moreover, the co-annihilation between $\tilde{\chi}_1^0$ and $\tilde{\chi}_1^\pm$ must be taken into account to obtain the reasonable DM relic density.
{{What is} %MDPI: We removed the bold. Please confirm this revision.
 more, it is noticed that our work indicates that SUSY processes with degenerate masses of $\tilde{\chi}_1^0$ and  $\tilde{\chi}_1^\pm$ within $10\%$ in the range from approximately 96~GeV to 240 GeV can be made at LHC Run 3 or HL-LHC to validate or disprove our model's~assumptions.
}

\vspace{6PT}

\acknowledgments{We thank Junjie Cao for helpful discussions.
This work is supported by the National Research Project Cultivation Foundation of Henan Normal University under Grant No. 2021PL10 and powered by the High-Performance Computing Center of Henan Normal University.}

\end{document}